\documentclass{Interspeech}

\interspeechcameraready

\title{Weakly Supervised Data Refinement and Flexible Sequence Compression for Efficient Thai LLM-based ASR}

\author[affiliation={1}]{Mingchen}{Shao}
\author[affiliation={1}]{Xinfa}{Zhu}
\author[affiliation={1}]{Chengyou}{Wang}
\author[affiliation={1}]{Bingshen}{Mu}
\author[affiliation={2}]{Hai}{Li}
\author[affiliation={2}]{Ying}{Yan}
\author[affiliation={2}]{Junhui}{Liu}
\author[affiliation={2}]{Danming}{Xie}
\author[affiliation={1,*}]{Lei}{Xie}

\affiliation{Audio, Speech and Language Processing Group (ASLP@NPU)}{\\Northwestern Polytechnical University}{China}
\affiliation{iQIYI}{Inc}{China}

\email{mcshao@mail.nwpu.edu.cn, lxie@nwpu.edu.cn\thanks{*Corresponding author.}}

\usepackage{comment}
\usepackage{graphicx}  
\usepackage{lscape}    
\usepackage{stfloats}
\usepackage{algorithm}
\usepackage{booktabs}   
\usepackage{array}      
\usepackage{amsmath}    
\usepackage{algpseudocode}
\usepackage{url}
\usepackage{xurl}

\begin{document}

\maketitle

\begin{abstract}
                                  

    Despite remarkable achievements, automatic speech recognition (ASR) in low-resource scenarios still faces two challenges: high-quality data scarcity and high computational demands. This paper proposes EThai-ASR, the first to apply large language models (LLMs) to Thai ASR and create an efficient LLM-based ASR system. EThai-ASR comprises a speech encoder, a connection module and a Thai LLM decoder. To address the data scarcity and obtain a powerful speech encoder, EThai-ASR introduces a self-evolving data refinement strategy to refine weak labels, yielding an enhanced speech encoder. Moreover, we propose a pluggable sequence compression module used in the connection module with three modes designed to reduce the sequence length, thus decreasing computational demands while maintaining decent performance. Extensive experiments demonstrate that EThai-ASR has achieved state-of-the-art accuracy in multiple datasets. We release our refined text transcripts to promote further research.\footnote{https://huggingface.co/datasets/mcshao/EThai-ASR}
    
\end{abstract}

\section{Introduction}
Recent advances in automatic speech recognition (ASR) have led to significant improvements in high-resource scenarios. However, in low-resource settings, limited annotated data and computational resources hinder performance, particularly in Thai ASR. Meanwhile, large language models (LLMs), such as GPT~\cite{gpt4} and LLama~\cite{2encoder}, have demonstrated impressive text understanding capabilities, which has spurred their increasing integration into ASR tasks. Recent approaches combine LLMs with speech encoders to leverage both linguistic knowledge and speech features, resulting in a general LLM-based ASR framework~\cite{qwen_audio,qwen2_audio,salmonn}. Nevertheless, the deployment of LLM-based ASR in low-resource scenarios like Thai ASR faces two main challenges: limited data availability and substantial computational demands~\cite{low1,low2}.

First, the scarcity of precisely annotated data is a significant challenge. 
Like many other low-resource languages in the world, Thai faces the difficulty of limited precisely annotated resources.
Only 500 hours of precisely annotated data are available in CommonVoice Thai~\cite{commonvoice}, compared to 680k hours in English LibriSpeech~\cite{libir}. 
The most available data for Thai comes from weak labels crawled on the web, which introduce label noise that can undermine ASR model training.  
For example, datasets like Gigaspeech2~\cite{gigaspeech2} and MSR-86k~\cite{msr} rely heavily on such weak labels. Although automated weak labels can expand the corpus, noisy labels present a bottleneck that hampers the ability to train ASR models with high accuracy~\cite{thaiasr1,thaiasr2,thaiasr3}.

Second, the high computational demands of LLMs severely restrict their application. 
Leveraging a large text dataset of Thai, Typhoon-2~\cite{typhoon,typhoon2} has developed a series of Thai LLMs that are ideally suited for developing Thai LLM-based ASR. 
Nevertheless, the high computational demands for training and inference continue to limit LLM-based ASR's practical applicability in low-resource scenarios. 
For instance, training models such as Seed-ASR~\cite{seedasr} may require up to 500k GPU hours which is unaffordable for most researchers~\cite{audiopalm,thaillm}.
A significant factor is the highly similar and redundant speech frames in the feature sequences produced by the speech encoder.
This redundancy results in substantial overhead for the LLM. 
However, conventional methods such as fixed downsampling and the CTC spike-based key frame extraction approach proposed by Skipformer ~\cite{skipformer} may inevitably discard essential speech features critical for the Thai language. 
Additionally, these ad hoc methods are architecture-dependent, often requiring the entire model to be trained from-scratch, complicating their adaptation to pretrained LLM-based ASR systems.

To address these challenges, we propose EThai-ASR for efficient and accurate Thai ASR, incorporating three key innovations. 
First, we introduce a self-evolving data refinement strategy that iteratively improves the quality of weak labels by leveraging a pretrained Zipformer~\cite{Zipformer} model and limited high-quality annotated data. 
Notably, we open-source the refined labels for Gigaspeech2~\cite{gigaspeech2} and MSR-86k~\cite{msr}, providing the community with 16k hours of refined speech data labels in Thai.   
Second, we propose an LLM-based ASR architecture that combines the enhanced Zipformer encoder obtained from the data iteration process with a Thai LLM, which is both rich acoustic and semantic.
This combination leads to state-of-the-art (SOTA) ASR model performance. 
Additionally, we introduce a two-stage training approach: in the first stage, the connection between the speech encoder and the LLM is trained to align the speech space and LLM's semantic space, while in the second stage, the LoRA components are finetuned to adapt to the ASR task.
Third, we propose a pluggable sequence compression module that dynamically removes redundant speech frames based on cosine similarity. 
This module substantially reduces computational overhead, achieving 1.5\(\times\) to 2.1\(\times\) acceleration while
incurring an accuracy degradation of less than 5\% without any modifications to the underlying model architecture. 

In summary, our contributions are threefold:
\begin{itemize}
    \item \textbf{Self-evolving data refinement.}  Iterative weak label correction via model feedback and open-sourcing refined labels for large-scale datasets.
    
    \item \textbf{LLM-based Thai ASR.} An integration of the enhanced Zipformer encoder with a Thai LLM achieving SOTA accuracy using a two-stage training strategy.
    
    \item \textbf{Pluggable sequence compression.} A pluggable sequence compression module that accelerates training and inference without modifying the LLM-based ASR model or severely sacrificing performance.
\end{itemize}

\begin{figure*}[htbp]  
    \centering
    \includegraphics[clip, trim=0cm 4.6cm 1.3cm 2cm, width=0.95\linewidth]{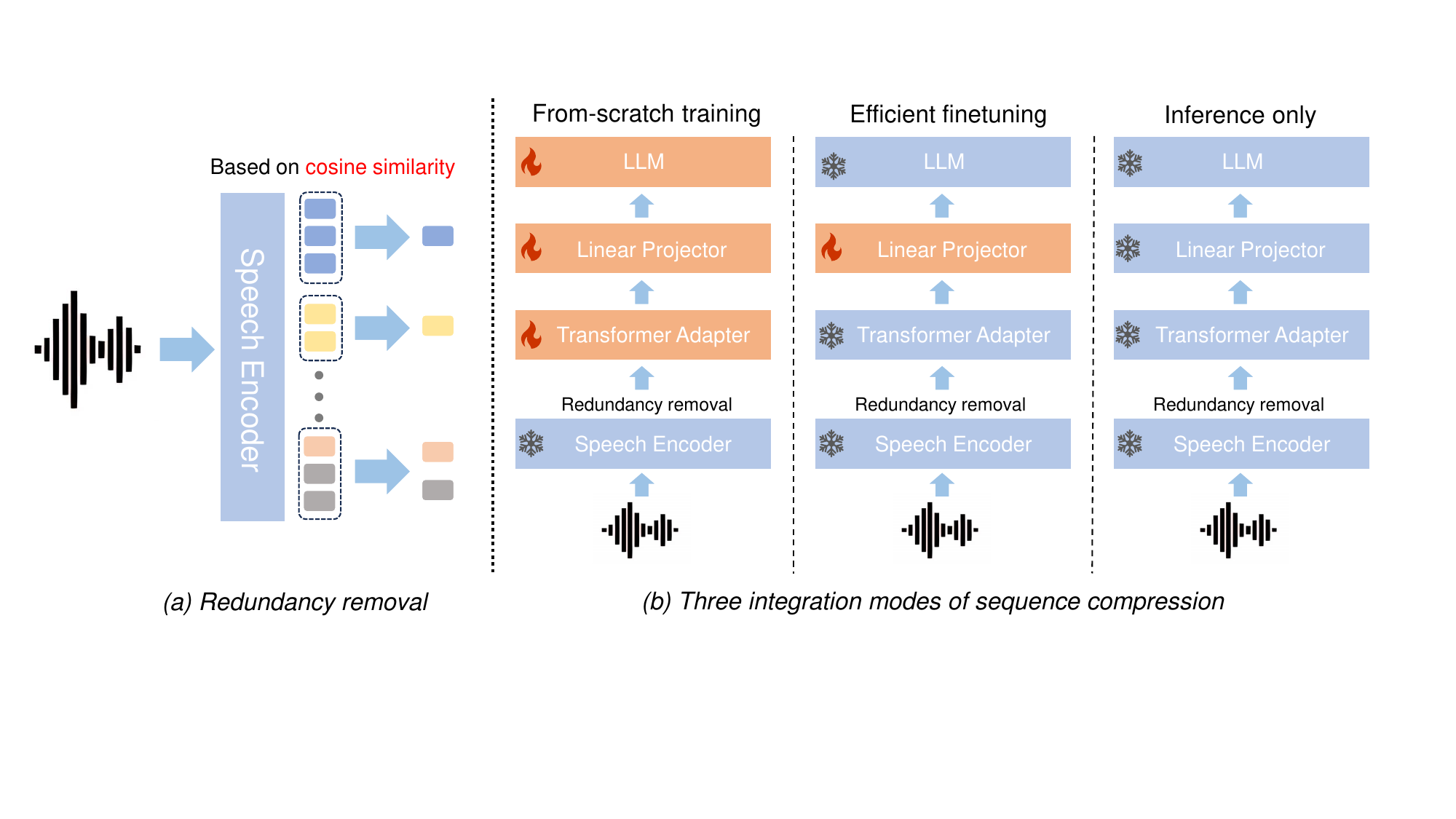}  
    \caption{\textbf{The architecture of EThai-ASR.} The left panel illustrates redundant speech frames are dynamically removed via cosine similarity to reduce sequence length before LLM processing. The right panel illustrates three integration modes: in from-scratch training, the Transformer adapter, linear projector, and LLM LoRA are finetuned after redundancy removal; efficient finetuning adjusts only the linear projector for comparable performance with reduced overhead; inference applies redundancy removal without any parameter training.}
    \label{fig:framework}
\end{figure*}

\section{Method}

EThai-ASR aims to leverage prior knowledge from text LLM for low-resource Thai ASR and address two challenges: data scarcity and high computational demands. To achieve this goal, EThai-ASR first introduces a self-evolving data refinement strategy to improve the quality of weakly labeled datasets. Subsequently, EThai-ASR is trained on the refined dataset, incorporating a speech encoder and a Thai LLM. Furthermore, a pluggable sequence compression is introduced to improve the efficiency and reduce computational demands.

\subsection{Self-evolving data refinement}

We propose a self-evolving data refinement~\cite{xinfa} illustrated by Algorithm 1 that iteratively enhances weak labels through model feedback. 
Process begins with a hybrid dataset $\mathcal{D} = \mathcal{D}_p \cup \mathcal{D}_w$, which comprises 500 hours of precise labels $\mathcal{D}_p$ and 16k hours of weak labels $\mathcal{D}_w$. 
First, our method finetunes the Zipformer-gigaspeech2 model~\cite{gigaspeech2}, using the hybrid dataset $\mathcal{D}$.
Besides, we use the Monsoon-whisper-medium-gigaspeech2 model\footnote{\url{https://huggingface.co/scb10x/monsoon-whisper-medium-gigaspeech2}}, a whisper-medium model finetuned on the Gigaspeech2 dataset, as an anchor for data selection.
Next, the Zipformer-gigaspeech2 model is employed to label the hybrid dataset, with the labels compared to the anchor labels to calculate the edit distance.
Suppose the edit distance between the labels is below a certain threshold. 
In that case, both models are considered to agree on the label, and the label produced by Zipformer is updated as the refined label~\cite{refine1,refine2,refine3}.
Otherwise, the sample is discarded.
Then, we use the hybrid dataset with the refined labels to continue training the Zipformer model, iterating in this way to complete the self-evolving data refinement.
This self-evolving data refinement process ultimately yields refined labels and a more performant Zipformer model, referred to as Zipformer-iter3 in this paper, which outperforms the original Zipformer-gigaspeech2 model after three rounds of data iteration.

\begin{algorithm}
\caption{Self-evolving label refinement}
\begin{algorithmic}[1]
\State \textbf{Input:} Hybrid dataset $\mathcal{D} = \mathcal{D}_p \cup \mathcal{D}_w$, Threshold $\tau$, Number of iterations $n$
\State \textbf{Output:} Refined label set $R$, Final model $M$
\State Initialize Zipformer-gigaspeech2 model $M$ with pre-trained weights
\State $R \gets \mathcal{D}$  \text{ // Initialize refined label set with hybrid dataset}
\For{$i \gets 1$ \textbf{to} $n$}
    \State Train model $M$ on $R$
    \State Generate predictions $\hat{y}$ by passing wav $x$ through $M$
    \State Compare predictions with anchor labels $y$ 
    \State Calculate edit distance between  $\hat{y}$ and $y$
    \State $R \gets \{(x, \hat{y}) \mid \text{edit distance}(\hat{y}, y) \leq \tau, (x, y) \in \mathcal{D}\}$
\EndFor
\State \textbf{Return:} Refined label set $R$, Final model $M$
\end{algorithmic}
\end{algorithm}

\subsection{LLM-based ASR architecture}
\label{sebsection:2.2}
To enhance the performance of the Thai ASR system, we integrate the Zipformer-iter3 encoder with a Thai LLM. 
The speech features extracted by the Zipformer-iter3 encoder are projected into the LLM's semantic space using a 4-layer Transformer adapter and a linear projector~\cite{transformer,qformer,projector}. 
This integration utilizes the advanced speech representations learned by the Zipformer-iter3 while also leveraging the contextual and linguistic knowledge inherent in the Thai LLM.
Additionally, we employ a two-stage training strategy. 
In the first stage, we unfreeze the Transformer adapter and linear projector to align the speech feature space with the LLM's semantic space. 
In the second stage, we further finetune the LoRA~\cite{lora,lora2} components of the LLM for optimal performance.

\subsection{Pluggable sequence compression}

In the LLM-based ASR outlined in Section~\ref{sebsection:2.2}, the encoder's output sequence contains many similar speech frames. 
These redundant speech frames lead to significant computational overhead when processed by the LLM.
To enhance the efficiency of the Thai ASR system in low-resource scenarios, we propose a pluggable sequence compression module that uses cosine similarity to remove redundant and highly similar speech frames from the Zipformer-iter3 encoder's output.
This method greatly compresses the sequence length that the LLM processes, reducing memory usage and running time, which are essential for training and inference. 
Moreover, the module can be flexibly applied in multiple modes, including from-scratch training, efficient fine-tuning, and inference only.

\subsubsection {Redundancy removal based on cosine similarity}
In speech recognition, the encoder's output often contains numerous redundant speech frames that contribute little to the final transcription~\cite{com2,com1}.
To address this issue, we compute the cosine similarity between adjacent frames and remove those that exceed a predefined similarity threshold. 
By computing frame-wise cosine similarity, our method enables adaptive compression.
Specifically, if the cosine similarity between two adjacent speech frames exceeds a threshold $\theta$, the frames are considered redundant, and the latter of them is removed. The cosine similarity between adjacent frames $\mathbf{x}_t$ and $\mathbf{x}_{t+1}$ is computed as follows:
\begin{align}
\text{CosineSim}(\mathbf{x}_t, \mathbf{x}_{t+1}) &= \frac{\mathbf{x}_t \cdot \mathbf{x}_{t+1}}{\lVert\mathbf{x}_t\rVert \lVert\mathbf{x}_{t+1}\rVert}
\label{equation:cosine_sim}
\end{align}

\subsubsection {Integration and application}
 As a pluggable solution, the proposed compression module is highly adaptable and supports three integration modes.

 \textbf{From-scratch training.} We train an LLM-based ASR model from-scratch that incorporates a pluggable sequence compression module to accelerate both training and inference.
 Specifically, we compare adjacent speech frames in the speech encoder's output and remove the later speech frame if its cosine similarity exceeds a preset threshold.
 In this mode, we use the two-stage strategy proposed in Section~\ref{sebsection:2.2}.
 This strategy reduces GPU memory consumption and time costs for both training and inference.

 \textbf{Efficient finetuning.}
 While removing redundant adjacent frames significantly reduces the input sequence length, this reduction in speech features may lead to misalignment between speech space and LLM's semantic space, as the LLM-based ASR was initially trained without compression.
 To mitigate this issue and rebuild the connection between the speech and LLM's semantic space, we employ a minimal amount of data to efficiently finetune the linear projector while keeping other components frozen, thereby enhancing the alignment between the two spaces \cite{visionzip,visionzip2}.
 In the efficient finetuning mode, only the linear projector layer with approximately 3M trainable parameters is updated, enabling seamless module integration into existing architectures.
 This efficient finetuning requires only a few hours of data and can be completed in 30 minutes using 8 Nvidia 3090 GPUs, making it both efficient and resource-saving.

 \textbf{Inference only. }
 Our proposed sequence compression method can also be applied during inference to accelerate processing. Feeding the compressed speech feature sequence directly into the LLM significantly reduces the computational overhead.

\section{Experiment}
\subsection{Experiment setup}

Our experiments employ a combination of weak labels and precise labels. Specifically, the weak labels data consist of Gigaspeech2 (13k hours) and MSR-86k (3k hours), while the precise labels dataset is CommonVoice Thai (500 hours).
We evaluate our models on three test sets: two in-domain sets (Gigaspeech2 Test and CommonVoice Thai Test) and one out-of-domain set (FLEUR Test~\cite{fleur}).
We compare our method with four baselines: Zipformer-gigaspeech2, Whisper-large-v2, Whisper-large-v3~\cite{whisper} and Monsoon-whisper-medium-gigaspeech2, the Monsoon-whisper-medium-gigaspeech2 is a Whisper-medium model finetuned on Gigaspeech2.
For our LLM-based ASR, we integrate the Zipformer-iter3 encoder with Thai LLM Typhoon2-Llama3.1-8B\footnote{\url{https://huggingface.co/scb10x/llama3.1-typhoon2-8b-instruct}} and Typhoon2-Llama3.2-3B\footnote{\url{https://huggingface.co/scb10x/llama3.2-typhoon2-3b-instruct}}.
Model performance is evaluated using CER for accuracy and the Speedup Ratio (SR) for acceleration. SR is calculated as follows:
\begin{equation}
\text{SR} = \frac{T_{\text{original}}}{T_{\text{accelerated}}}
\end{equation}
where \( T_{\text{original}} \) is the time for the unaccelerated model, and \( T_{\text{accelerated}} \) is the time for the accelerated model. A higher SR indicates better acceleration.

\subsection{Evaluation of self-evolving data refinement}

By applying our self-evolving refinement, we iteratively filter out samples with an edit distance greater than 10 over three iters. 
As shown in Table~\ref{tab:self_evolving_results}, this method results in continuous reductions in CER, achieving absolute improvements of 0.38\% to 0.69\%, which correspond to relative improvements ranging from 3.05\% to 16.63\% when compared to the Zipformer-gigaspeech2 baseline.
The results indicate that our self-evolving data refinement method refines labels and mitigates data scarcity in low-resource scenarios.
It refines the labels and yields a better Thai ASR model, Zipformer-iter3, surpassing the Zipformer-gigaspeech2 baseline.

\begin{table}[ht]
  \caption{\textbf{CER (\%) performance through self-evolving data refinement across test sets over three iterations.} “Giga2 Test” indicates the Gigaspeech2 test set, “CV Test” denotes the CommonVoice test set, and “Iter. 0” corresponds to the baseline model.} 
  \label{tab:self_evolving_results}
  \centering
  \begin{tabular}{ l c c c c }
    \toprule
    \textbf{\parbox{1.3cm}{ Iter.}} & 
    \textbf{\parbox{1cm}{\centering Giga2 Test}} & 
    \textbf{\parbox{1cm}{\centering CV Test}} & 
    \textbf{\parbox{1cm}{\centering FLEUR Test}} & 
    \textbf{\parbox{1cm}{\centering Average CER}} \\
    \midrule
    0 (Baseline) & 12.46 & 4.15 & 10.54 & 9.05 \\
    1            & 12.31 & 3.45 & 10.94 & 8.90 \\
    2            & 12.14 & 3.52 & 10.73 & 8.79 \\
    3            & \textbf{12.08} & \textbf{3.46} & \textbf{10.21} & \textbf{8.58} \\

    \bottomrule
  \end{tabular}
\end{table}

\subsection{Evaluation of LLM-baed ASR}

Our LLM-based ASR integrates the Zipformer-iter3 encoder with Typhoon2-Llama3.1-8B via a 4-layer Transformer adapter and a linear projector. 
The model is trained using a two-stage training strategy. 
In the second stage, the LLM's LoRA layer is finetuned with a LoRA rank of 32 and a LoRA alpha set to 16. 
Both stages use a learning rate of 5e-4.

Table~\ref{tab:llm_performance} presents the CER comparisons across different models. 
The results demonstrate that our model significantly outperforms various baseline models. 
Specifically, when compared to Zipformer-iter3, our model achieves a 3.4\% relative reduction in CER on the Gigaspeech2 test set from 12.08\% to 11.67\% and shows a 42.9\% improvement on the FLEUR test set relative to Whisper-large-v2 from 15.50\% to 9.36\%.
By using the Zipformer-iter3 encoder and a powerful LLM as the decoder, we combine speech and semantic information to achieve SOTA performance in Thai ASR.
\begin{table}[ht]
  \caption{\textbf{LLM-based ASR CER (\%) performance  across models.} Test sets include Gigaspeech2 (Giga2 Test), CommonVoice (CV Test), and FLEUR Test. Our proposed model integrates Zipformer-iter3 with Typhoon2-Llama3.1-8B.}
  \label{tab:llm_performance}
  \centering
  \begin{tabular}{ l c c c c }
    \toprule
    \textbf{\parbox{0.8cm}{ Model }} & 
    \textbf{\parbox{0.8cm}{\centering Giga2 Test}} & 
    \textbf{\parbox{0.8cm}{\centering CV Test}} & 
    \textbf{\parbox{0.8cm}{\centering FLEUR Test}} \\
    \midrule
    Monsoon-whisper-medium & 14.15 & 6.92 & 11.69 \\
    Whisper-large-v3            & 20.44 & 6.02 & 11.55 \\
    Whisper-large-v2            & 22.47  & 8.79 & 15.50 \\
    Zipformer-gigaspeech2            & 12.46 & 4.15 & 10.54 \\
    Zipformer-iter3            & 12.08 & 3.46 & 10.21 \\
    EThai-ASR (our proposed)            & \textbf{11.67} & \textbf{3.11} & \textbf{9.36} \\
    \midrule
  \end{tabular}
\end{table}
\subsection{Analysis of pluggable sequence compression}
In this section, we evaluate the performance of the proposed sequence compression module. We conduct two experiments to demonstrate the effectiveness and flexibility of the module. 
\begin{figure}[h] \raggedright \includegraphics[width=0.49\textwidth]{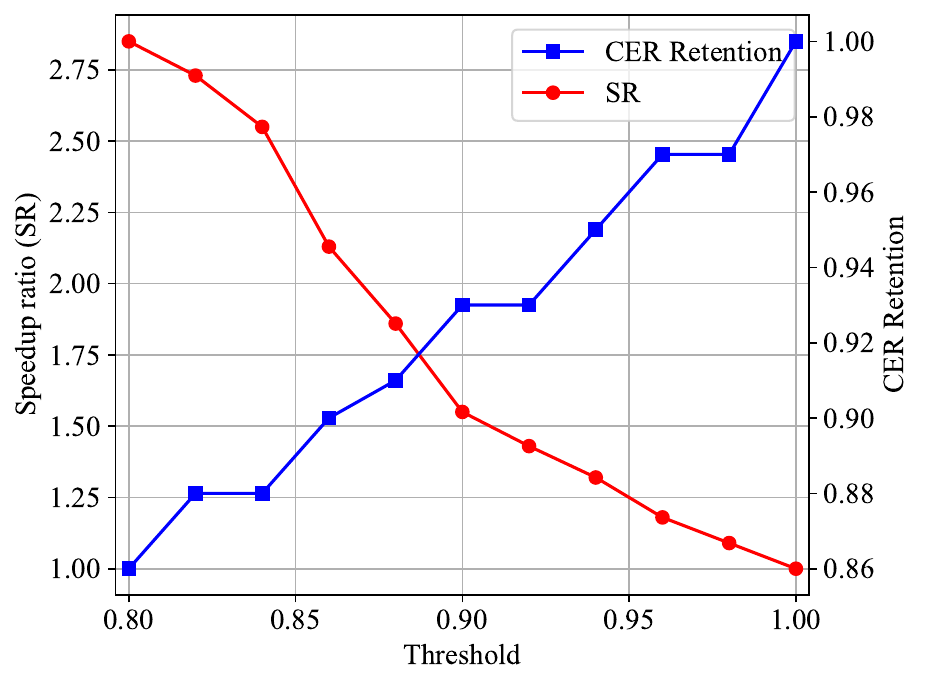} \caption{\textbf{Impact of cosine similarity thresholds on SR and CER retention.} The red line represents the variation in SR, while the blue line illustrates the changes in CER Retention.} \label{fig:threshold_tradeoff} \end{figure}
\subsubsection{Threshold-performance-acceleration tradeoff}

We evaluate the impact of the cosine similarity threshold $\theta$ on both the SR and CER retention.
Figure~\ref{fig:threshold_tradeoff} reveals that as the threshold decreases, the proportion of sequence compression increases, leading to more significant acceleration, while the CER retention degrades. A balance between acceleration and CER retention is achieved when the threshold reaches 0.887.
This shows that our sequence compression method can regulate the compression ratio using a threshold, providing a balance between SR and CER retention by adjusting the threshold.

\subsubsection{Compatibility across multiple LLM backbones}

To further validate the generality of our compression module, we evaluate its performance on different LLM backbones under three operational modes: from-scratch training, finetuning, and inference.
For from-scratch training and finetuning, we set the threshold to 0.87 to maintain a 25\% compression ratio, enabling a comparison with traditional 4\(\times\) convolutional downsampling. For inference, we increase the threshold to 0.95 to maintain a 50\% compression ratio. Table~\ref{tab:cross_backbone} demonstrates that our compression method achieves comparable performance to traditional downsampling across three LLM backbones while retaining over 95\% of the performance compared to models without sequence compression. Furthermore, the efficient finetuning mode achieves comparable performance to from-scratch training and traditional compression methods while requiring fewer training parameters and less data, and it offers the flexibility of a pluggable solution. In the direct inference mode, model performance degrades significantly as the compression ratio increases.

\begin{table}[H]
  \caption{\textbf{Cross-backbone CER (\%) performance under different operational modes.} Comparison includes full sequence processing (Full token), 4$\times$ convolutional downsampling (4$\times$Conv), and our method at $\theta=0.87/0.95$. Inference results show threshold sensitivity.}
  \label{tab:cross_backbone}
  \centering
  \begin{tabular}{m{1.5cm} l c c}  
    \toprule
    \textbf{Model} & \textbf{Mode} & 
    \textbf{threshold} & \textbf{CER} \\
    \midrule
    Typhoon-7B & 
    \begin{tabular}[c]{@{}l@{}} 
    Full token\\ 4\(\times\)Conv \\ From-scratch\\ Efficient finetuning\\ Inference only
    \end{tabular} & 
    \begin{tabular}[c]{@{}c@{}} 
    - \\ - \\ 0.87 \\ 0.87 \\ 0.87/0.95 
    \end{tabular} & 
    \begin{tabular}[c]{@{}c@{}} 
    16.87 \\ 16.97 \\ 17.24 \\ 17.37 \\ 37.17/27.31 
    \end{tabular} \\
    
    \midrule
    Typhoon2-Llama3.2-3B & 
    \begin{tabular}[c]{@{}l@{}} 
    Full token\\ 4\(\times\)Conv \\ From-scratch\\ Efficient finetuning\\ Inference only
    \end{tabular} & 
    \begin{tabular}[c]{@{}c@{}} 
    - \\ - \\ 0.87 \\ 0.87 \\ 0.87/0.95 
    \end{tabular} & 
    \begin{tabular}[c]{@{}c@{}} 
    14.76 \\ 14.89 \\ 14.87 \\ 14.96 \\ 32.50/14.95 
    \end{tabular} \\
    
    \midrule
    Typhoon2-Llama3.1-8B & 
    \begin{tabular}[c]{@{}l@{}} 
    Full token\\ 4\(\times\)Conv \\ From-scratch\\ Efficient finetuning\\ Inference only
    \end{tabular} & 
    \begin{tabular}[c]{@{}c@{}} 
    - \\ - \\ 0.87 \\ 0.87 \\ 0.87/0.95 
    \end{tabular} & 
    \begin{tabular}[c]{@{}c@{}} 
    11.67 \\ 11.88 \\ 11.83 \\ 11.92 \\ 28.97/14.10 
    \end{tabular} \\
    \bottomrule
  \end{tabular}
\end{table}

\section{Conclusion}

In this work, we propose EThai-ASR for low-resource scenarios. We introduce a self-evolving data refinement strategy that improves weak labels, enhances recognition accuracy by projecting speech features into a Thai LLM's semantic space using a two-stage training strategy, and reduces computational overhead through threshold-aware sequence compression.
Experimental results show significant improvement in ASR accuracy and acceleration of training and inference, demonstrating the effectiveness of EThai-ASR in low-resource scenarios.

\newpage
    
     

\bibliographystyle{IEEEtran}
\bibliography{mybib}

\end{document}